# Magnetic Structure and Properties of the S = 5/2 Triangular Antiferromagnet α-NaFeO$_2$

T. McQueen,[1] Q. Huang,[2] J. W. Lynn,[2] R. F. Berger,[1] T. Klimczuk,[3,4] B. G. Ueland,[5] P. Schiffer,[5] and R. J. Cava[1]

[1]*Department of Chemistry, Princeton University, Princeton, NJ 08544*
[2]*NIST Center for Neutron Research, National Institute of Standards and Technology, Gaithersburg, MD 20899*
[3]*Los Alamos National Laboratory, Los Alamos, NM 87545*
[4]*Faculty of Applied Physics and Mathematics, Gdansk University of Technology, Narutowicza 11/12, 80-952 Gdansk, Poland*
[5]*Department of Physics and Materials Research Institute, Pennsylvania State University, University Park, PA 16802*

## ABSTRACT

The magnetic properties of α-NaFeO$_2$ are studied by neutron diffraction and magnetization measurements. An ordered phase with spins aligned along the $\vec{b}_{hex}$ axis exists at low temperatures ($T < 4K$). At intermediate temperatures ($4K < T < 11K$), the system passes through an incommensurate ordered phase before transforming into a short range ordered state at higher temperatures that persists up to at least $50K$. Although the short range ordering does not persist to room temperature according to neutron diffraction, the magnetic susceptibility does not follow Curie-Weiss behavior, even up to $320K$. This rich magnetic behavior can be understood qualitatively as a competition between different magnetic exchange interactions that are similar in magnitude. The delicate balance between these interactions makes α-NaFeO$_2$ a candidate for more detailed theoretical work to understand magnetic behavior in frustrated magnetic systems.

## INTRODUCTION

Sodium ferrite (α-NaFeO$_2$) is the prototype structure for a large class of NaMO$_2$ layered oxides (see Figure 1(a)). Compounds in this class have the trigonal space group R-3m and consist of two dimensional MO$_2$ layers of edge sharing MO$_6$ octahedra separated by sodium ions. Adjacent MO$_2$ layers are offset laterally to create a three-layer structure. The sodium atoms occupy the octahedral holes between these layers. Due to the triangular lattice of M atoms within a plane and the relatively large separation between layers, a variety of magnetic and electrical properties result, depending on the identity of the M atom and the relative strengths of the various magnetic exchange interactions (representative interactions are drawn in Figure 1(b)). In compounds where interplane ($J_{ip}$) interactions dominate, spin glass behavior (as in LiNiO$_2$)[1] and antiferromagnetism (as in NaNiO$_2$ or LiNiO$_2$)[1,2] are common. In systems where $J_{ip}$ is comparatively weak, 2-D magnetic behavior is often observed (as in NaCrO$_2$).[3] The recent discovery of superconductivity in the hydrated two-layer variant of Na$_x$CoO$_2$, Na$_{0.3}$CoO$_2$·1.3H$_2$O,[4] and its α-NaFeO$_2$-type three-layered form[5,6] has renewed interest in these types of materials.

Sodium ferrite is a reddish-brown electrical insulator and is antiferromagnetic.[7-10] There are several previous reports on the magnetic structure of α-NaFeO$_2$. Tomkowicz and B. Van Laar determined the magnetic structure at $4.2K$ based on neutron diffraction[7], but there were small unexplained peaks in the data. Allain and Parette found a second magnetic structure to explain the extra peaks but did not elaborate on the details of the structure or comment on the diffuse scattering that they observed at low angles.[10] Here we report the detailed study of the magnetic transitions in α-NaFeO$_2$ based on powder neutron diffraction and magnetization measurements. Surprisingly, sodium ferrite shows a delicate balance between competing magnetic interactions that gives rise to short range and long range ordered magnetic phases with magnetic susceptibility that is not fit to any simple model up to $320K$.

## EXPERIMENTAL METHODS

α-NaFeO$_2$ was synthesized as previously reported.[9] Requisite amounts of α-Fe$_2$O$_3$ (Johnson Matthey, 99.99%) and 5% excess Na$_2$O$_2$ (Alfa Aesar, 93%) were mixed in a glove box, placed in an alumina crucible, and fired in air at $773K$ for 2 hours. The powder was then reground, pressed into a pellet, and annealed in air at $973K$ for 48 hours. The resulting reddish-brown powder was confirmed to be single phase α-NaFeO$_2$. Neutron diffraction patterns were taken at the NIST center for neutron research using

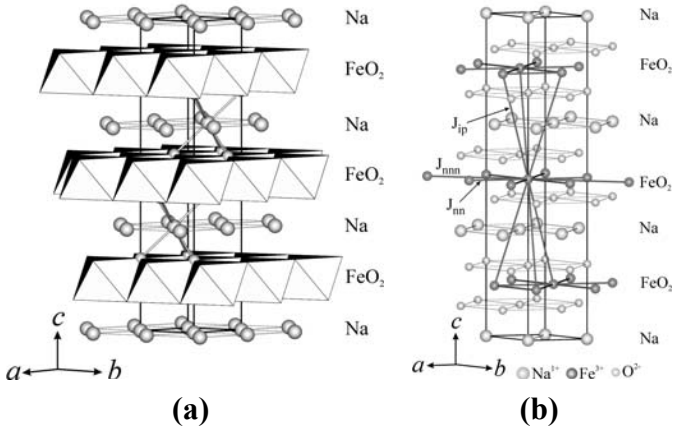

**(a)**          **(b)**

**Figure 1**. (a) The α-NaFeO$_2$ structure type consists of hexagonal FeO$_2$ layers separated by planes of Na atoms. These layers are offset laterally to create a "three-layer" structure. (b) The important magnetic exchange interactions within a metal layer are between nearest neighbors ($J_{nn}$) and next nearest neighbors ($J_{nnn}$). The nearest interaction between planes ($J_{ip}$) also plays a critical role in the observed magnetic structures.

the BT-1 powder diffractometer and the BT-9 triple axis spectrometer. Roughly $5g$ of finely ground α-NaFeO$_2$ were packed in a vanadium canister and mounted on a closed-cycle refrigerator. Low angle (3-28° two theta) powder patterns were taken at temperatures between $3.5K$ and $50K$ using a (filtered) pyrolytic graphite (002) monochromator, $\lambda = 2.3591(2)\text{Å}$, on BT-9. High resolution patterns were taken at $298K$ and $4K$ using a Cu (311) monochromator with a 90° take-off angle, $\lambda = 1.5404(2)\text{Å}$, and an in-pile collimation of 15′ on BT-1. Additional patterns suitable for magnetic structure refinement were taken at $20K$, $15K$, $11K$, $10K$, $9K$, $8K$, $7K$, $6K$, $5.5K$, $5K$, $4.5K$, $4.0K$ and $3.5K$ using a Ge (311) monochromator with a 75° take-off angle, $\lambda = 2.0789(2)\text{Å}$, and an in-pile collimation of 15′ on BT-1. All BT-1 data were collected over the two theta range 3–168° with a step size of 0.05°. FullProf 2000 and WinPlotR[11] were used for structural and magnetic Rietveld refinements. Scattering lengths (in *fm*) employed during the refinements were 3.63, 9.45, and 5.803 for Na, Fe, and O respectively. The magnetic form factor $\langle j_0 \rangle$ coefficients tabulated by P.J. Brown were used for the magnetic ion, Fe$^{3+}$.[12] The rhombohedral setting of space group R-3m (R-3mR) was used for all refinements. Magnetization measurements were performed in Quantum Design PPMS ($T = 1.8 - 320K$, $\mu_0 H = 1T$) and MPMS ($T = 1.8K - 50K$, $\mu_0 H = 1 - 5T$) magnetometers.

## CRYSTALLOGRAPHIC AND MAGNETIC STRUCTURE

The reported crystallographic structure at $298K$ for α-NaFeO$_2$ is confirmed.[13] Crystallographic parameters in the space group R-3m, including atomic positions and displacement parameters, from the refinement of neutron powder data at $298K$ ($\chi^2 = 1.57$) are shown in Table 1. These values agree well with those previously reported and confirm the three layer structure at room temperature.

**Table 1**. Structural parameters for α-NaFeO$_2$ at 298 K obtained using the space group R-3mR. The atomic positions are: **Na**: 1a (0 0 0), **Fe**: 1b (½ ½ ½), and **O**: 2c (x x x). All sites are fully occupied.

| R-3m | Rhombohedral Setting | | | Hexagonal Setting | | | |
|---|---|---|---|---|---|---|---|
| | a | 5.64184(4) Å | | a | 3.02511(4) Å | | |
| | α | 31.1016(3)° | | c | 16.09410(7) Å | | |
| Atom | x | y | z | x | y | z | $B_{iso}$ (Å$^2$) |
| Na | 0 | 0 | 0 | 0 | 0 | 0 | 0.848(20) |
| Fe | 0.5 | 0.5 | 0.5 | 0 | 0 | 0.5 | 0.512(11) |
| O | 0.23344(3) | 0.23344(3) | 0.23344(3) | 0 | 0 | 0.23344(3) | 0.563(12) |
| $\chi^2 = 1.57$ $R_{wp} = 7.48$ $R_p = 8.94$ | | | | | | | |

The trigonal structure is maintained at the lowest temperature studied, $3.5K$. At low temperatures, additional diffraction peaks from magnetic scattering are observed. Figure 2 shows a series of neutron powder diffraction ($\lambda = 2.0789\text{Å}$) patterns over the two theta range 3-80° taken at $3.5K$, $5K$, $8K$, and $20K$. At $3.5K$, extra peaks corresponding to an ordered magnetic phase (magnetic II) are visible and marked with triangles. By $5K$, the magnetic II phase coexists with a second long range ordered magnetic phase (magnetic I), whose peaks are marked with circles. By $8K$, only the magnetic I and nuclear peaks remain, but there is weak diffuse scattering observed around $2\theta \approx 11°$. At $20K$, all sharp peaks are fit by the nuclear model, but the broad diffuse scattering peak at low angle is still prominent. This broad peak is still weakly present at $50K$ but not visible at room temperature (Figure 2).

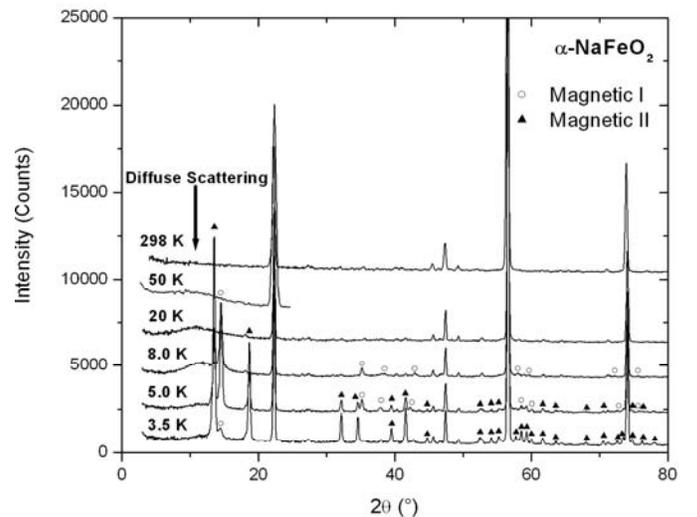

**Figure 2**. Neutron diffraction shows that α-NaFeO$_2$ displays several distinct magnetic phases below room temperature: a weak diffuse scattering peak from short range order; an incommensurate long range magnetic ordering, magnetic I, marked with circles; and a commensurate magnetic phase, magnetic II, marked with triangles. To facilitate direct comparisons, the two theta values for the data at 50 K and 298 K were adjusted for the differences in wavelengths used during collection.

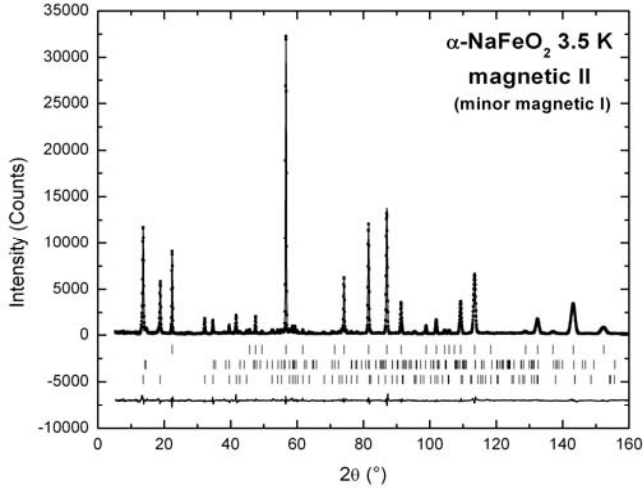

**Figure 3**. Fit of neutron diffraction data at 3.5 $K$ to the nuclear (first set of tick marks), magnetic I (second set of tick marks) and magnetic II (third set of tick marks) contributions. The diffuse scattering peak was subtracted prior to Rietveld refinement.

The magnetic ordering in the low temperature magnetic II phase is commensurate with the primitive rhombohedral lattice. This phase is fully developed at $3.5K$, with only minor magnetic I peaks (< 5%) remaining. Full profile refinement of these data confirms the magnetic structure reported by Z. Tomkowicz and B. Van Laar for α-NaFeO$_2$ at $4.2K$.[7] The fit to the $3.5K$ data is shown in Figure 3 (the minor diffuse scattering peak has been pre-fit and removed). The agreement is very good ($\chi^2 = 3.47$, $R_{mag} = 10.46$) with no unindexed peaks or systematic trends in the residuals. An adequate fit is obtained only if the minority magnetic I phase is included in the refinement.

In the magnetic II phase, all the magnetic moments on individual Fe$^{3+}$ ions are equal in magnitude and aligned either parallel or antiparallel with the $\vec{b}_{hex}$ axis. The spins on adjacent iron sites within an FeO$_2$ layer are ordered along the $\vec{b}_{hex}$ direction to form lines of parallel spins. These lines are ordered along the $\vec{a}_{hex}$ direction to form stripes of two-up and two-down spins ( ↑↑↓↓ ), as shown in the left of Figure 4(a). The overall ordering within each layer is therefore antiferromagnetic. These planes of antiferromagnetic ordering are stacked in the $\vec{c}_{hex}$ direction such that each spin has 4 antiparallel and 2 parallel neighbors in adjacent planes, as shown in the right of Figure 4(a).

At $8K$, the magnetic I phase is well-developed, but the magnetic II phase is not present. Consequently, the $8K$ diffraction data are the most appropriate for refinement of the first magnetic phase. The $2\theta$ and d-spacing values for magnetic I reflections, determined by comparison of the data at $8K$ to that at $20K$, are given in Table 2. These reflections are not easily indexed as a (small) supercell of the nuclear model, so it was assumed that the phase is incommensurate. Following established procedures,[11, 14-17] propagation vectors that index these peaks as satellites of the primitive rhombohedral lattice (nuclear model) were determined. Peak assignments of the observed reflections are given in Table 2. The propagation vectors found are all cyclic permutations of each other. Although it is impossible to distinguish between these possibilities from the powder data, it is not necessary to do so as they are symmetry equivalent and yield identical magnetic structures. Thus for convenience the vector $\vec{k}_{rhom} = <0.380, 0.500, 0.618>$ will be used. Using this propagation vector, the Fourier coefficients of the magnetic moments were determined by Rietveld refinement of the $8K$ data. The Fourier coefficients correspond to moments pointing along the crystallographic direction $\hat{\mu}_{rhom} = <-0.15, 1.14, -0.81>$. Figure 5 shows the $8K$ data fit to this model. The agreement is excellent ($\chi^2 = 2.50$, $R_{mag} = 16.62$). There are no unindexed peaks or systematic trends in the residuals. In the present model, the direction of the spins is not restricted to be perpendicular to $\vec{k}_{rhom}$, and, as a result, fits better than the model by Allain and Parette[10] ($R_{mag} = 16.62$ vs. $R_{mag} = 19.49$).

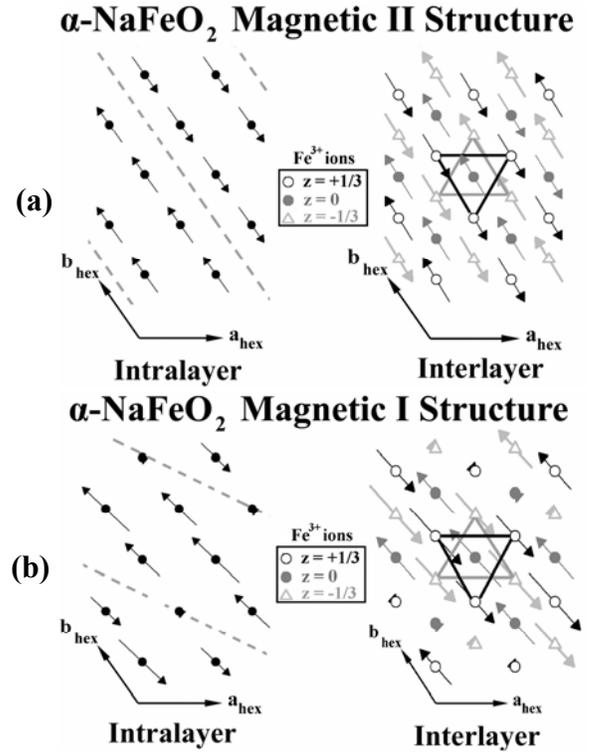

**Figure 4**. Sodium ferrite displays two long-range-ordered magnetic phases. Part (a) shows the ordering of the magnetic II phase within an iron plane (left) and between planes (right). Part (b) shows the ordering of the magnetic I phase within an iron plane (left) and between planes (right). Planes of iron atoms are identified by their z-value in the hexagonal setting of R-3m.

**Table 2**. The magnetic I peaks that were identified by comparison of neutron data at 20 $K$ and 8 $K$. The d-spacings of the magnetic I peaks at 5 $K$ are also listed, along with the shift in each peak from 8 $K$ to 5 $K$. The reflection assignments for each peak are listed. The propagation vector and magnetic moment direction are also given; the magnetic moment $\hat{\mu}_{rhom}$ points almost perpendicular to $\vec{k}_{rhom}$ (88°).

| 8 $K$ | | 5 $K$ | | |
|---|---|---|---|---|
| 2 θ (°) | d-spacing (Å) | d-spacing (Å) | δ (5 $K$ – 8 $K$) | (h k l)+k$_{rhom}$ Assignments |
| 14.534 | 8.1799 | 8.1799 | 0 | (0,0,0)+,(-1,-1,-1)+ |
| 35.174 | 3.4338 | 3.4338 | 0 | (1,1,1)+,(-2,-2,-2)+ |
| 38.51 | 3.1468 | 3.1935 | 0.0467 | (0,-1,-1)+,(0,0,-1)+ |
| 42.91 | 2.8376 | 2.8765 | 0.0389 | (1,0,0)+,(-1,-1,-2)+ |
| 58.03 | 2.1408 | 2.1272 | -0.0136 | (2,1,1)+,(-2,-2,-3)+ |
| 59.673 | 2.0871 | 2.0748 | -0.0123 | (0,0,1)+,(-2,-1,-1)+ |
| 72.233 | 1.7621 | 1.7544 | -0.0077 | (1,1,2)+,(-3,-2,-2)+ |
| 75.657 | 1.6936 | 1.6936 | 0 | (1,0,-1)+,(0,-1,-2)+ |
| **Propagation Vector and Moment Direction at 8 $K$** | | | | |
| Propagation Vector | krhom = <.38,.50,.62> | | khex = <-.12,-.12,1.50> | |
| Moment Direction | μrhom =<-.15,1.14,-.81> | | μhex = <-.21,.87,.06> | |

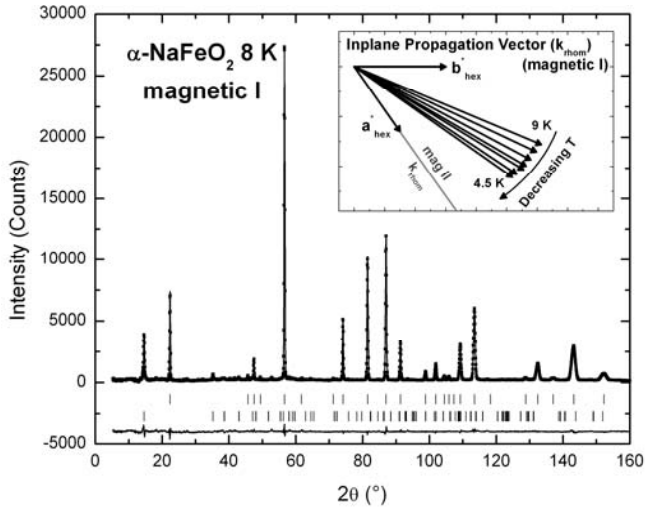

**Figure 5**. Fit of neutron diffraction data at 8 *K* to the nuclear (first set of tick marks) and magnetic I (second set of tick marks) contributions. The diffuse scattering peak was subtracted prior to Rietveld refinement. The inset shows the intraplane component of the propagation vector describing magnetic I, highlighting the systematic change with temperature.

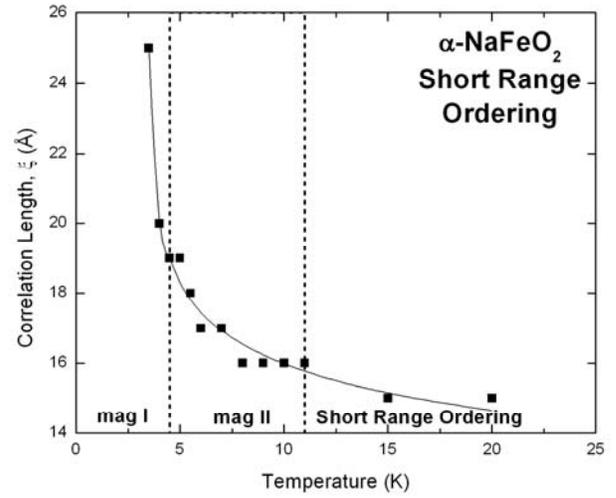

**Figure 6**. The width of the diffuse scattering peak at low angles ($2\theta \approx 11°$) can be used to extract the average correlation length for the short range ordering. As expected, the correlation length increases as temperature decreases. The curve is drawn to guide the eye.

In this magnetic I phase, the spins lie almost within the iron layers, although there is a non-negligible component perpendicular to the layers at $8K$ (~6% of the total moment). Within a single iron layer, the spins have a periodic modulation forming alternating antiparallel stripes that are incommensurate with the underlying lattice. The projection of spins within a single Fe layer is drawn in the left of Figure 4(b). The dashed lines highlight that the alternating stripes of spins are separated by lines of iron atoms with little ordered moment. The overall ordering within each plane, including the component perpendicular to the layers, is antiferrromagnetic. The stripes in adjacent layers are stacked so that the majority of interactions are antiferromagnetic; the right of Figure 4(b) shows this arrangement of spins. For all sites that have appreciable ordered moments, the nearest neighbors in adjacent planes, marked by dashed lines, are all aligned antiparallel. It is only in the border regions of small ordered moments that all spins are not aligned antiferromagnetically between layers.

There is a significant temperature dependence of the ordering in the magnetic I phase. This can be seen in Figure 2 between the $8K$ and $5K$ data. While the magnetic reflections at 14.5° and 35.2° are unchanged, the reflections at 38.5° and 42.9° at $8K$ shift to 37.9° and 42.3° respectively at $5K$. A full list of the magnetic I peaks at $5K$ compared to $8K$ is given in Table 2. The reflection corresponding to the satellite of the (000) reflection does not shift with temperature. This implies that the propagation vector describing the incommensurate modulation changes in direction but not magnitude as the temperature decreases. This is observed in the refinements: the inset in Figure 5 shows the direction of $\vec{k}_{rhom}$ within a layer versus temperature. As temperature goes down, the propagation vector rotates away from the $\vec{b}_{hex}$ axis and the repetition length of spins increases along that axis. This is interesting because the magnetic II phase has an infinite repetition length of spins along the $\vec{b}_{hex}$ axis, so the incommensurate magnetic phase is transitioning to the magnetic II ordering.

The third kind of magnetic scattering observed is a diffuse peak centered at $2\theta \approx 11°$. The broad nature of the diffuse scattering peak is characteristic of short range ordering (SRO), but a detailed analysis is not possible with the present data. As such, a single pseudo-Voigt peak was used to fit this scattering in all subsequent refinements. The dependence of the SRO (diffuse scattering) on temperature can be seen in the correlation length, $\xi$, which can be extracted from the peak width.[11, 18] As expected, the correlation length increases as temperature decreases, from $\xi = 15 \mathring{A}$ at $20K$ to $\xi = 25 \mathring{A}$ at $3.5K$ (Figure 6). Since the direction of the propagation vector describing the SRO responsible for the diffuse scattering is not known, the integrated area of the diffuse peak can only be used to establish an upper limit on the magnetic moment contribution from the phase.[11, 18] Figure 7 shows the dependence of this integrated area as well as the moments of the magnetic I and II phases versus temperature. Interestingly, the SRO persists through the magnetic I transition, and it is not until the low temperature magnetic II phase develops that the diffuse scattering intensity substantially decreases. This suggests that the transition to the magnetic I long range ordered state on cooling is fundamentally different in nature than the transition from the magnetic I to the magnetic II state.

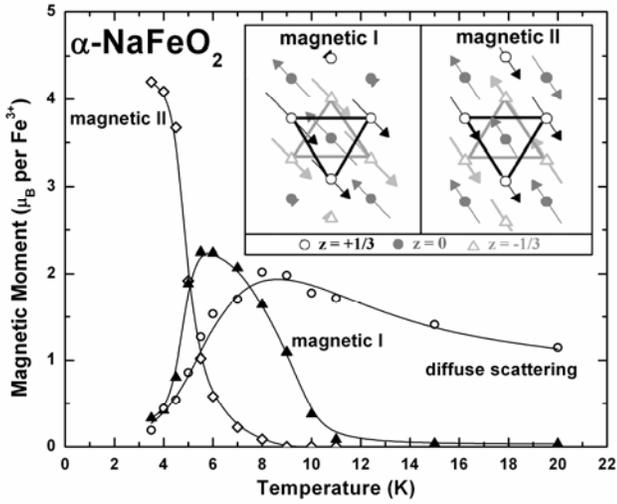

**Figure 7**. Sodium ferrite displays two long-range-ordered magnetic phases as well as short range magnetic correlations. The moment per $Fe^{3+}$ was extracted from fits of the diffuse scattering and Rietveld refinement at each temperature. The inset shows the ordering of the moments in the magnetic I and magnetic II phases as determined by neutron refinements.

The magnetic phases observed in the neutron powder diffraction patterns agree with the measured magnetization of α-NaFeO$_2$. Figure 8(a) shows the magnetic susceptibility of a polycrystalline sample at $1T$ from $320K$ to $1.8K$. The susceptibility shows that α-NaFeO$_2$ is antiferromagnetic with a Néel temperature of $T_N = 11K$. This coincides with the appearance of the magnetic I phase from the neutron refinements. The abrupt change in slope at $5K$, shown in the Figure 8(a) inset, coincides with the transition from the magnetic I to the magnetic II phase. The broad tail on the transition above the Néel temperature is consistent with the short range magnetic ordering observed in the neutron data. Figure 8(b) shows the inverse susceptibility at $1T$ without (filled squares) and with (open circles) the subtraction of a $\chi_0$ of $-0.0164 \frac{emu}{molFe^{3+} \cdot Oe}$. Although correction with a $\chi_0$ does make the high temperature region nearly linear, as shown in the inset, the value of $\chi_0$ is unusually large. Additionally, the calculated effective magnetic moment with this $\chi_0$ is $11.94 \mu_B / Fe^{3+}$, significantly larger than what is expected for high-spin iron (III), $5.9 \mu_B / Fe^{3+}$. Furthermore, the first derivative still has an upward curvature, even after subtraction (see inset). Consequently, the Curie-Weiss law is not a good description of the high temperature behavior of α-NaFeO$_2$. This is probably due to the competing exchange interactions which frustrate the spins, producing short range magnetic fluctuations that persist to quite high temperatures as observed in the neutron data.

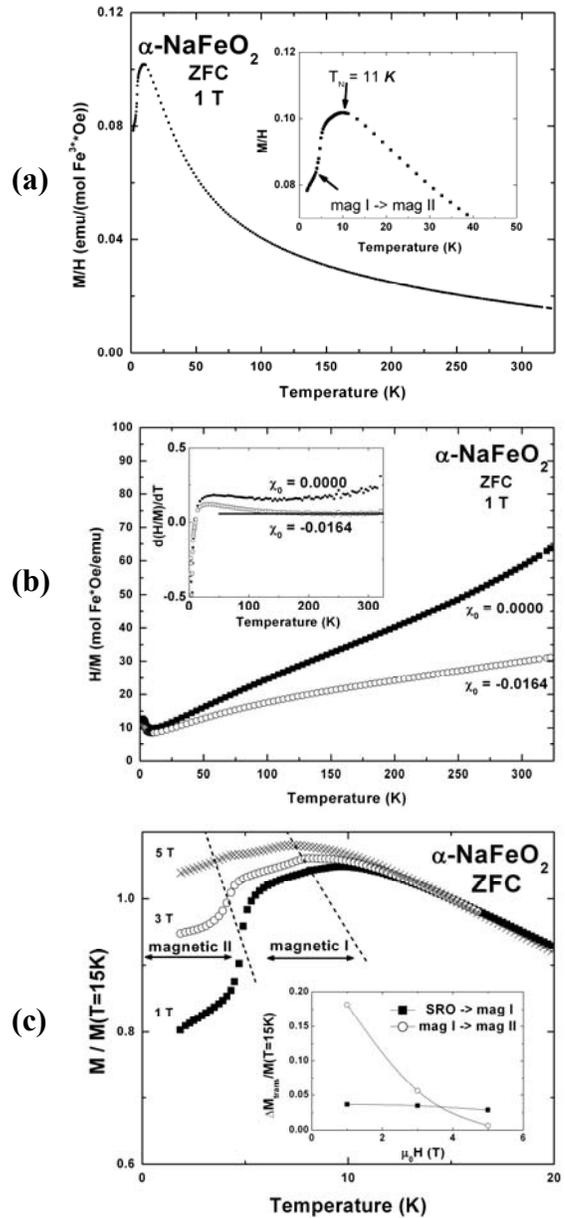

**Figure 8**. (a) The magnetic transitions in α-NaFeO$_2$ can be seen in the zero field cooled magnetic susceptibility data at $\mu_0 H = 1$ $T$. The Néel point coincides with the appearance of the magnetic I phase in the neutron data. A change in slope in the susceptibility around $5$ $K$ is concomitant with the transition from the magnetic I to the magnetic II phase. (b) α-NaFeO$_2$ does not follow Curie-Weiss behavior up to $320$ $K$, even with a $\chi_0$ term. (c) Magnetization curves normalized to the values at T = $15$ $K$ show the suppression of the magnetic I to magnetic II phase transition.

The transition from the magnetic I to magnetic II phase is strongly dependent on the applied magnetic field. Figure 8(c) shows the normalized magnetization from $1.8K$ to $20K$ at $1T$, $3T$, and $5T$. The larger applied fields slightly reduce the temperature of both magnetic transitions; the magnetic I to magnetic II transition is reduced from $5K$ to $4.2K$, and the short range ordering (SRO) to magnetic I transition is lowered to $8K$ from $11K$. The magnetic I to magnetic II transition is almost entirely suppressed in the $5T$ field, but the magnitude of the SRO to magnetic I transition only shows a slight decrease, as shown in the inset. This is consistent with the results derived from neutron diffraction, i.e., that the transition to the magnetic I long range ordered state on cooling is fundamentally different in nature than the transition from the magnetic I to the magnetic II state.

## DISCUSSION

The successive magnetic phases in α-NaFeO$_2$ can be understood qualitatively as a competition between magnetic interactions with three types of magnetic neighbors. Figure 1(b) shows the magnetic exchange interactions that are assumed to be important: a ferromagnetic nearest neighbor in-plane (NN) interaction $J_{nn}$, an antiferromagnetic next nearest neighbor in-plane (NNN) interaction $J_{nnn}$, and an antiferromagnetic nearest neighbor interplane (IP) interaction $J_{ip}$. In the magnetic I phase, the IP alignment is antiparallel, and the NN alignment is mainly parallel. The NNN ordering is a mix of parallel and antiparallel orientations. This suggests that the two nearest neighbor interactions, $J_{nn}$ and $J_{ip}$, are stronger than the NNN interaction, $J_{nnn}$, although $J_{nnn}$ cannot be neglected. The NN interactions are ferromagnetic, but the NNN interactions are non-negligible and antiferromagnetic. The NNN coupling frustrates the spin system and prevents each layer from achieving a long-range ferromagnetic state. Thus the magnetic I state represents a compromise between ferromagnetic NN and antiferromagnetic NNN interactions.

At temperatures below $8K$, the system transitions to the magnetic II phase. The magnetic II phase differs from the magnetic I phase most in the alignment of the spins and the interplane ordering. The spins in magnetic II are aligned along the $\vec{b}_{hex}$ axis with all spins equal in magnitude and only the sign of the moment varying. Significantly, the interlayer alignments are no longer solely antiparallel: the IP ordering between layers is 4 antiparallel / 2 parallel. Additionally, the NN orientations are 4 parallel / 2 antiparallel and the NNN alignments are 4 antiparallel / 2 parallel. This partial satisfaction of all interactions suggests that all three effective exchange interactions are approximately balanced $(-J_{nn} \approx J_{nnn} \approx J_{ip})$.

This change in the relative strengths of exchange interactions explains why the magnetic I and magnetic II phases are distinct rather than displaying a continuous transition from one to the other. While the propagation vector of the magnetic I phase approaches that of the magnetic II phase within an iron layer as temperature decreases (Figure 5 inset), consistent with the antiferromagnetic NNN interaction becoming more important relative to the ferromagnetic NN one, the propagation vector along the $\vec{c}_{hex}$ axis stays fixed at $\frac{3}{2}$. This reflects the relative strength of the antiferromagnetic IP exchange, which prevents the system from adopting more complex spin arrangements that permit the propagation vector along $\vec{c}_{hex}$ to vary. However, as soon as the magnitude of $J_{nnn}$ outweighs that of $J_{ip}$, the strict antiferromagnetic ordering between planes becomes unstable and the system adopts a different interplane ordering. The new ordering is that of the magnetic II phase, described by a propagation vector of 1 along the $\vec{c}_{hex}$ axis.

Within this model, the short range ordering above T$_N$ can then be explained as correlations between spins due to either the ferromagnetic NN or antiferromagnetic IP interactions, or both, where long range order has not set in due to thermal excitation of the spins. Unfortunately, neither the neutron scattering nor magnetic susceptibility data is sufficient to differentiate between these possibilities. The fact that the magnetic susceptibility does not follow Curie-Weiss behavior up to the highest temperatures studied is probably due to the competing exchange interactions and consequent frustration, which results in short range spin correlations: it does not fit simple Heisenberg (isotropic, single axis anisotropic, single plane anisotropic) or Ising models.

It is useful to compare the magnetic structures observed in α-NaFeO$_2$ to two structurally similar compounds, CuFeO$_2$ (Fe$^{3+}$) and MnBr$_2$ (Mn$^{2+}$). CuFeO$_2$ has the delafossite structure type, which differs with α-NaFeO$_2$ only in the positions of the oxygen atoms. The low temperature magnetic phase in CuFeO$_2$ is related to the magnetic II phase but has strictly alternating stripes (↑↓) rather than two-up two down stripes (↑↑↓↓) within a layer.[19] More distinctly, the moments in CuFeO$_2$ are aligned parallel to the $\vec{c}_{hex}$ axis, and the ordering schemes are not the same, presumably due to the different oxygen positions in the delafossite structure type.

Conversely, MnBr$_2$, which has the same metal and anion arrangment as α-NaFeO$_2$ (Mn$^{2+}$ is isoelectronic with Fe$^{3+}$) but without sodium atoms between layers, has the same magnetic ordering within the plane as the magnetic II phase (↑↑↓↓). However, the adjacent layers are aligned differently with respect to each other so that there are 3 antiparallel / 3 parallel alignments between layers for every spin. This is probably a consequence of stronger coupling between layers in MnBr$_2$ due to the lack of a separating sodium layer (where the next-nearest-neighbor interaction between planes is also important).[20, 21]

## CONCLUSIONS

α-NaFeO$_2$ shows a series of successive magnetic phase transitions, determined by powder neutron diffraction and magnetization measurements. Above $11K$, short-range order is observed to persist to quite high temperatures. At lower temperatures the system displays two different long-range ordered phases. Magnetic I is antiferromagnetic and incommensurate with the underlying lattice. Magnetic II is also antiferromagnetic but commensurate with the lattice, forming stripes of ferromagnetic spins within the iron layers (↑↑↓↓). The structure of the magnetic I and magnetic II phases can be understood assuming that there are three important magnetic interactions that are similar in magnitude – the nearest neighbor interactions within an iron layer and between iron layers, and the next nearest neighbor interactions within a layer. Detailed theoretical calculations of the magnetic structure of α-NaFeO$_2$ would be of interest to confirm this qualitative model for the spin structures observed.

## ACKNOWLEDGEMENTS

T. McQueen gratefully acknowledges support by the National Science Foundation Graduate Research Fellowship Program. This work was done under the NSF MRSEC program, grant NSF-DMR06-20234m and grant NSF-DMR-0353610. Certain commercial materials and equipment are identified in this report to describe the subject adequately. Such identification does not imply recommendation or endorsement by the NIST, nor does it imply that the materials and equipment identified is necessarily the best available for the purpose.

## REFERENCES

[1] C. M. Julien, A. Ait-Salah, A. Mauger, et al., Ionics **12**, 21 (2006).
[2] C. Darie, P. Bordet, S. de Brion, et al., European Physical Journal B **43**, 159 (2005).
[3] A. Olariu, P. Mendels, F. Bert, et al., Physical Review Letters **97** (2006).
[4] K. Takada, H. Sakurai, E. Takayama-Muromachi, et al., Nature **422**, 53 (2003).
[5] R. E. Schaak, T. Klimczuk, M. L. Foo, et al., Nature **424**, 527 (2003).
[6] M. L. Foo, T. Klimczuk, L. Li, et al., Solid State Communications **133**, 407 (2005).
[7] Z. Tomkowicz and B. Vanlaar, Physica Status Solidi A: Applied Research **23**, 683 (1974).
[8] T. Ichida, T. Shinjo, Y. Bando, et al., Journal of the Physical Society of Japan **29**, 795 (1970).
[9] Y. Takeda, J. Akagi, A. Edagawa, et al., Materials Research Bulletin **15**, 1167 (1980).
[10] Y. Allain and G. Parette, Comptes Rendus Hebdomadaires Des Seances De L Academie Des Sciences Serie B **281**, 293 (1975).
[11] J. Rodriguez-Carvajal, Physica B **192**, 55 (1993).
[12] P. J. Brown, in *International Tables for Crystallography*, edited by A. J. C. Wilson, Vol. C, p. 391.
[13] M. S. Goldsztaub, Bulletin de la Societe Francaise de Mineralogie **58**, 6 (1935).
[14] D. E. Cox, IEEE Transactions on Magnetics **8**, 161 (1972).
[15] Y. A. Izyumov and V. E. Naish, Journal of Magnetism and Magnetic Materials **12**, 239 (1979).
[16] W. Sikora, F. Bialas, and L. Pytlik, Journal of Applied Crystallography **37**, 1015 (2004).
[17] C. Wilkinson, G. Lautenschlager, R. Hock, et al., Journal of Applied Crystallography **24**, 365 (1991).
[18] M. A. Krivoglaz, *X-Ray and Neutron Diffraction in Nonideal Crystals* (Springer-Verlag, Berlin Heidelberg, 1996).
[19] M. Mekata, N. Yaguchi, T. Takagi, et al., Journal of the Physical Society of Japan **62**, 4474 (1993).
[20] T. Sato, H. Kadowaki, H. Masuda, et al., Journal of the Physical Society of Japan **63**, 4583 (1994).
[21] T. Sato, H. Kadowaki, and K. Iio, Physica B **213**, 224 (1995).